\documentclass[11pt,twoside]{article}
\usepackage{asp2014}

\aspSuppressVolSlug
\resetcounters

\begin{document}

\title{Update of the China-VO AstroCloud}
\author{Chenzhou Cui $^1$, Ce Yu$^2$, Jian Xiao$^2$, Boliang He$^1$, Changhua Li$^1$, Dongwei Fan$^1$, Chuanjun Wang$^3$, Zhi Hong$^2$, Shanshan Li$^1$, Linying Mi$^1$, Wanghui Wan$^{1,4}$, Zihuang Cao$^1$, Jiawei Wang$^1$, Shucheng Yin$^2$, Yufeng Fan$^3$, Jianguo Wang$^3$, Sisi Yang$^1$,Yin Ling$^1$,Hailong Zhang$^5$,Junyi Chen$^3$,Liang Liu$^6$, Xiao Chen$^7$
\affil{$^1$National Astronomical Observatories, Chinese Academy of Sciences (CAS), 20A Datun Road, Beijing 100012, China; \email{ccz@nao.cas.cn}}
\affil{$^2$Tianjin University, 92 Weijin Road, Tianjin 300072, China; \email{xiaojian@tju.edu.cn}}
\affil{$^3$Yunnan Astronomical Observatory, CAS, P.0.Box110, Kunming 650011, China; \email{wcj@ynao.ac.cn}}
\affil{$^4$Wuhan Science and Technology Museum, 104 Zhaojiatiao Road, Wuhan 430010, China; \email{wanwh@nao.cas.cn}}
\affil{$^5$Xinjiang Astronomical Observatory, CAS, 150 Science 1-Street, Urumqi, Xinjiang 830011, China; \email{zhanghailong@xao.ac.cn}}
\affil{$^6$Purple Mountain Observatory, 2 West Beijing Road, Nanjing 210008, China; \email{liangliu@pmo.ac.cn}}
\affil{$^7$Shanghai Astronomical Observatory, 80 Nandan Road, Shanghai 200030, China ; \email{cx@shao.ac.cn}}
}

\paperauthor{Chenzhou Cui}{ccz@nao.cas.cn}{}{National Astronomical Observatories, Chinese Academy of Sciences (CAS)}{Center of Information and Computing}{Beijing}{Beijing}{100012}{China}
\paperauthor{Ce Yu}{yuce@tju.edu.cn}{}{Tianjin University}{School Of Computer Science And Technology}{Tianjin}{Tianjin}{300072}{China}
\paperauthor{Jian Xiao}{xiaojian@tju.edu.cn}{}{Tianjin University}{School Of Computer Science And Technology}{Tianjin}{Tianjin}{300072}{China}
\paperauthor{Boliang He}{hebl@nao.cas.cn}{0000-0002-3244-7312}{National Astronomical Observatories, Chinese Academy of Sciences (CAS)}{Center of Information and Computing}{Beijing}{Beijing}{100012}{China}
\paperauthor{Changhua Li}{lich@nao.cas.cn}{}{National Astronomical Observatories, Chinese Academy of Sciences (CAS)}{Center of Information and Computing}{Beijing}{Beijing}{100012}{China}
\paperauthor{Dongwei Fan}{fandongwei@nao.cas.cn}{}{National Astronomical Observatories, Chinese Academy of Sciences (CAS)}{Center of Information and Computing}{Beijing}{Beijing}{100012}{China}

\begin{abstract}
As the cyber-infrastructure for Astronomical research from Chinese Virtual Observatory (China-VO) project, AstroCloud has been archived solid progresses during the last one year. Proposal management system and data access system are redesigned. Several new sub-systems are developed, including China-VO PaperData, AstroCloud Statics and Public channel. More data sets and application environments are integrated into the platform. LAMOST DR1, the largest astronomical spectrum archive was released to the public using the platform. The latest progresses will be introduced.
\end{abstract}

\section{Introduction}
AstroCloud\citep{Cui_adassxxiv} is a cyber-infrastructure for astronomy research initiated by Chinese Virtual Observatory (China-VO) under funding support from NDRC (National Development and Reform commission) and CAS (Chinese Academy of Sciences). Tasks such as proposal management, data archiving, data quality control, data release and open access, Cloud based data processing and analyzing, are integrated into the physical distributed platform. It acts as a full lifecycle management system and gateway for astronomical data and telescopes.

\section{New feature}
Since its lunch on May 15, 2014, many new features and remarkable improvements have been made to the platform. The current system consists of five channels, i.e. Observation, Data, Tools, Cloud and Public.
\articlefigure{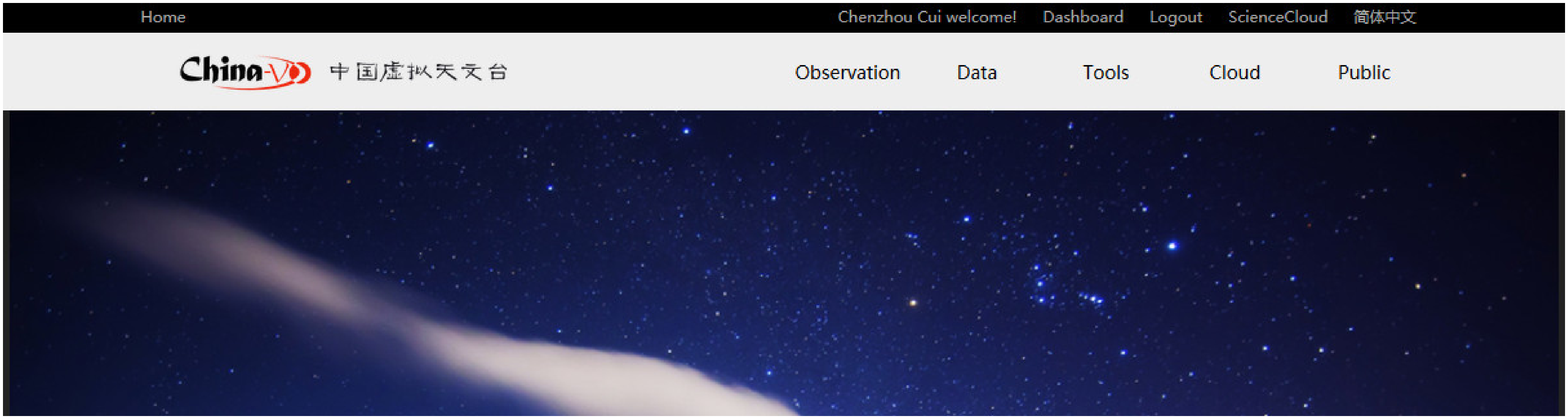}{ex_fig1}{Channels of AstroCloud}

Proposal submission sub-system\citep{P8-3_Xiao_adassxxiv} was re-designed to provide flexible management features for telescope managers. Configuration parameters for back-end instrument, proposal submission, peer-review, and observation time allocation can be changed by telescope managers easily.

\articlefigure{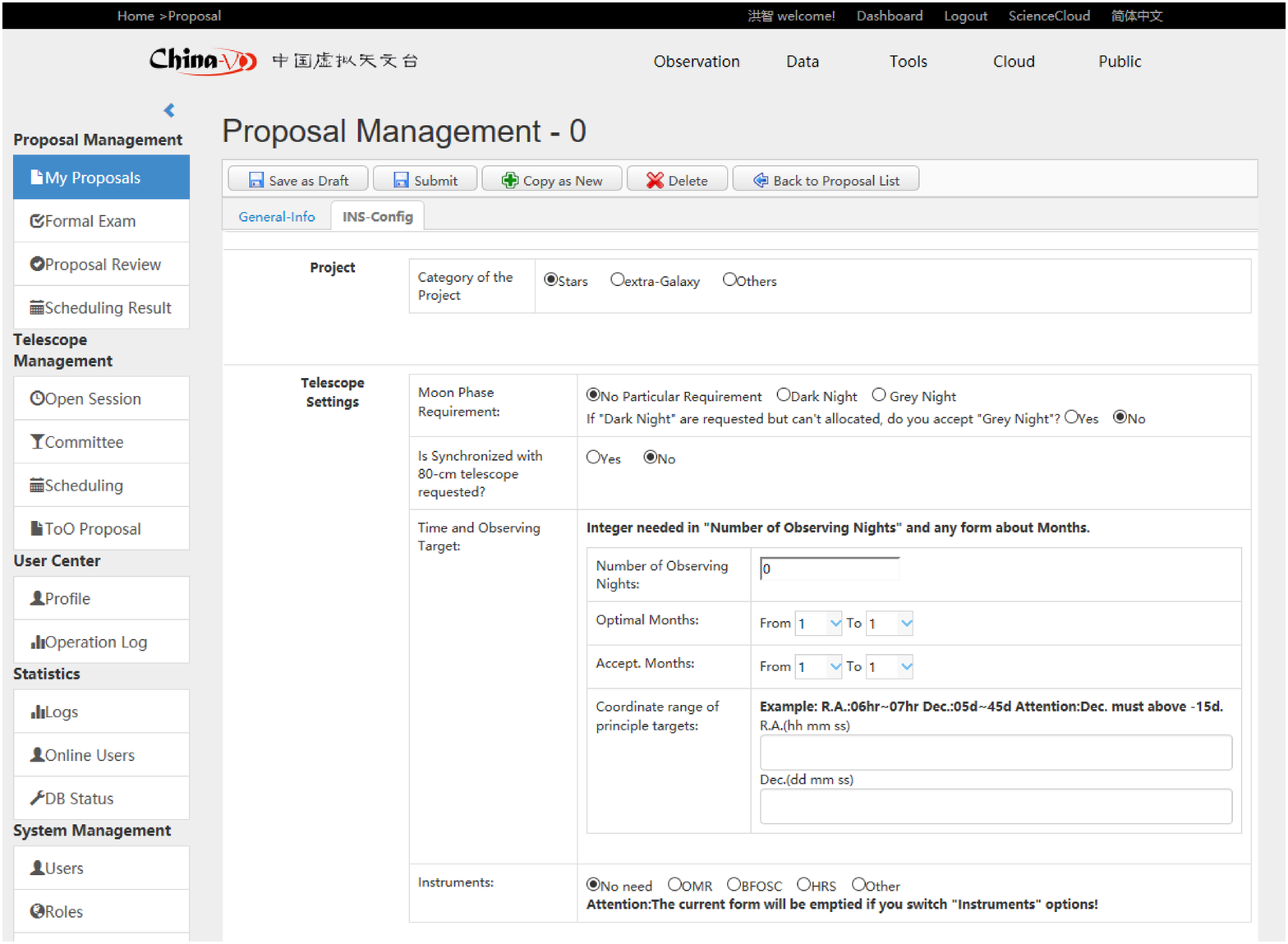}{ex_fig1}{Proposal Management System}

Data query and access sub-system \citep{P1-3_Fan_adassxxiv}was improved for better performance and interoperability. Tens of datasets hosted at the platform can be queried and cross identified through a uniform interface. LAMOST DR1\citep{2015RAA....15.1095L}, the largest astronomical spectrum dataset with 2.2M spectra was open to the public in March 2015 through the system.

\articlefigure{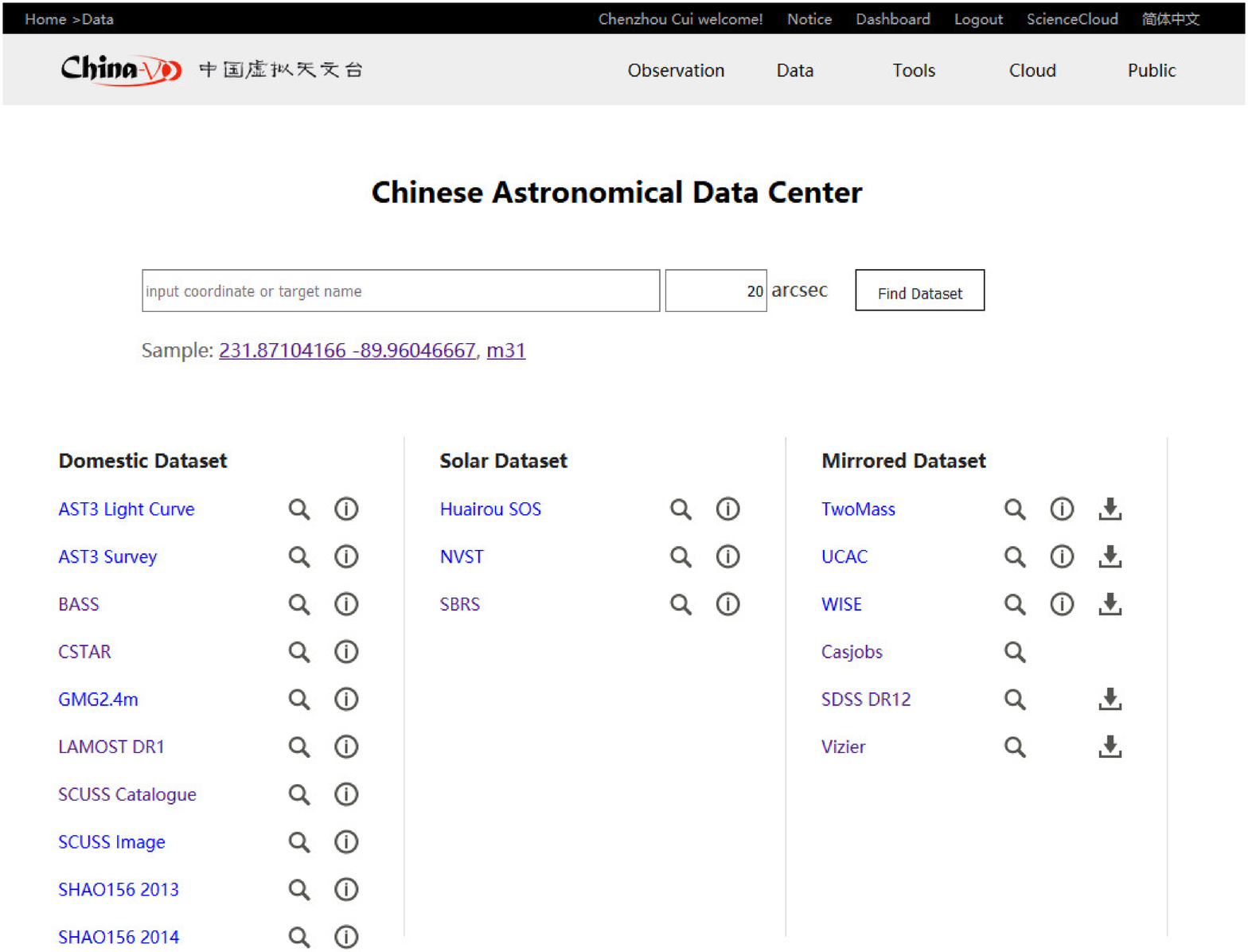}{ex_fig1}{Data Query and Access System}

A new dashboard is designed to give a user fast access points of frequently used resources and services, and summary information of the system and the user.

Acquisition, Reduction and Analysis of Multi-wavelength Astronomical Data (MADARA) is a Cloud Computing\citep{P1-2_Li_adassxxiv} based teaching and research environment for astronomical lectures and graduate students. Common used software packages, for example IRAF, DS9, CASA, HEASOFT, SSW, IDL, Python, to process and analyze multi-wavelength observation data are pre-installed. A virtual machine instance with these packages can be initiated and ready for using in only few minutes.

China-VO Paper Data Repository, provides long-term storage and open access service for user’s paper data, which includes but not limited tables, figures, pictures, movies, source codes, models, software packages mentioned in his scientific papers. A permanent but user specified URL will be provided for each item.

Public channel is a new one especially developed for the public and amateur astronomers in the last year. Video streams provide live images taken from video cameras at different observatories. Gallery is a collection of beautiful pictures taken by AstroCloud users. China-VO Special is a collection of China-VO hosted services, for example Astronomical Dictionary and WWT Beijing Community.

On July 29, 2015, Popular Supernova Project was lunched. It is the first astronomical citizen science project in China as a joint venture between China-VO and Xingming Amateur Astronomical Observatory. In the morning of Sep. 12, a supernova candidate was discovered by a 10-year old pupil. Inspired by the news, number of the registered users of AstroCloud platform raised violently to 105K by the end of September. Two supernova candidates discovered by public users in October 2015 have been confirmed by professional observations.

In additional to the above user facing channels, several crucial functions are provided by the backend platform. Two examples are given here.

CSTNET passport, the combination of an email address and a password provided by China Science \& Technology Network (CSTNET) that you use to sign in to supported services. If you don't have a CSTNET passport, you can sign up for free at any time.

Usage Statistics, give out important statistic data about the platform interested by users and administrators. For example, general weblog results and platform running status including online users, registered users, login in numbers, etc. Submitted proposals for telescopes. Archived astronomical observation datasets and their latest progress\citep{P5-1_He_adassxxiv}; Running status and number of virtual machine instances at each Cloud computing node.

\acknowledgements This paper is funded by National Natural Science Foundation of China (11503051, U1231108, U1531246, U1531115), Ministry of Science and Technology of China (2012FY120500), Chinese Academy of Sciences (XXH12503-05-05). Data resources are supported by Chinese Astronomical Data Center.

\bibliography{P023}  

\begin{thebibliography}{}
\expandafter\ifx\csname natexlab\endcsname\relax\def\natexlab#1{#1}\fi
\expandafter\ifx\csname url\endcsname\relax
  \def\url#1{\texttt{#1}}\fi
\expandafter\ifx\csname urlprefix\endcsname\relax\def\urlprefix{URL }\fi
\providecommand{\eprint}[2][]{\url{#2}}

\bibitem[{{Cui} et~al.(2015){Cui}, {Yu}, {Xiao}, {He}, {Li}, {Fan}, {Wang},
  {Hong}, {Li}, {Mi}, {Wan}, {Cao}, {Wang}, {Yin}, {Fan}, {Wang}, \&
  {Yang}}]{Cui_adassxxiv}
{Cui}, C., {Yu}, C., {Xiao}, J., {He}, B., {Li}, C., {Fan}, D., {Wang}, C.,
  {Hong}, Z., {Li}, S., {Mi}, L., {Wan}, W., {Cao}, Z., {Wang}, J., {Yin}, S.,
  {Fan}, Y., {Wang}, J., \& {Yang}, S. 2015, in ADASS XXIV, edited by A.~R.
  Taylor, \& E.~Rosolowsky (San Francisco: ASP), vol. 495 of ASP Conf. Ser.,
  469

\bibitem[{{Fan} et~al.(2015){Fan}, {He}, {Xiao}, {Li}, {Li}, {Cui}, {Yu},
  {Hong}, {Yin}, {Wang}, {Cao}, {Fan}, {Mi}, {Wan}, \&
  {Wang}}]{P1-3_Fan_adassxxiv}
{Fan}, D., {He}, B., {Xiao}, J., {Li}, S., {Li}, C., {Cui}, C., {Yu}, C.,
  {Hong}, Z., {Yin}, S., {Wang}, C., {Cao}, Z., {Fan}, Y., {Mi}, L., {Wan}, W.,
  \& {Wang}, J. 2015, in ADASS XXIV, edited by A.~R. Taylor, \& E.~Rosolowsky
  (San Francisco: ASP), vol. 495 of ASP Conf. Ser., 477

\bibitem[{{He} et~al.(2015){He}, {Cui}, {Fan}, {Li}, {Xiao}, {Yu}, {Wang},
  {Cao}, {Chen}, {Yi}, {Li}, {Mi}, \& {Yang}}]{P5-1_He_adassxxiv}
{He}, B., {Cui}, C., {Fan}, D., {Li}, C., {Xiao}, J., {Yu}, C., {Wang}, C.,
  {Cao}, Z., {Chen}, J., {Yi}, W., {Li}, S., {Mi}, L., \& {Yang}, S. 2015, in
  ADASS XXIV, edited by A.~R. Taylor, \& E.~Rosolowsky (San Francisco: ASP),
  vol. 495 of ASP Conf. Ser., 483

\bibitem[{{Li} et~al.(2015){Li}, {Wang}, {Cui}, {He}, {Fan}, {Yang}, {Chen},
  {Zhang}, {Yu}, {Xiao}, {Li}, {Mi}, {Wang}, {Cao}, {Fan}, {Hong}, {Wan},
  {Wang}, \& {Yin}}]{P1-2_Li_adassxxiv}
{Li}, C., {Wang}, J., {Cui}, C., {He}, B., {Fan}, D., {Yang}, Y., {Chen}, Y.,
  {Zhang}, H., {Yu}, C., {Xiao}, J., {Li}, S., {Mi}, L., {Wang}, C., {Cao}, Z.,
  {Fan}, Y., {Hong}, Z., {Wan}, W., {Wang}, J., \& {Yin}, S. 2015, in ADASS
  XXIV, edited by A.~R. Taylor, \& E.~Rosolowsky (San Francisco: ASP), vol. 495
  of ASP Conf. Ser., 487

\bibitem[{{Luo} et~al.(2015){Luo}, {Zhao}, \& {Zhao}}]{2015RAA....15.1095L}
{Luo}, A.-L., {Zhao}, Y.-H., \& {Zhao}, G. 2015, Research in Astronomy and
  Astrophysics, 15, 1095

\bibitem[{{Xiao} et~al.(2015){Xiao}, {Yu}, {Cui}, {He}, {Li}, {Fan}, {Hong},
  {Yin}, {Wang}, {Cao}, {Fan}, {Li}, {Mi}, {Wan}, {Wang}, \&
  {Zhang}}]{P8-3_Xiao_adassxxiv}
{Xiao}, J., {Yu}, C., {Cui}, C., {He}, B., {Li}, C., {Fan}, D., {Hong}, Z.,
  {Yin}, S., {Wang}, C., {Cao}, Z., {Fan}, Y., {Li}, S., {Mi}, L., {Wan}, W.,
  {Wang}, J., \& {Zhang}, H. 2015, in ADASS XXIV, edited by A.~R. Taylor, \&
  E.~Rosolowsky (San Francisco: ASP), vol. 495 of ASP Conf. Ser., 473

\end{thebibliography}

\end{document}